\documentclass[%
notitlepage,
reprint,
superscriptaddress,
amsmath,amssymb,
aps,
]{revtex4-2}

\usepackage{xcolor}
\usepackage{lipsum}
\usepackage{graphicx}
\usepackage{dcolumn}
\usepackage{bm}
\usepackage{hyperref}

\usepackage{epstopdf}

\newcommand{\bra}[1]{\langle #1 \vert}
\newcommand{\ket}[1]{\vert #1 \rangle}
\usepackage{ulem}

\usepackage{xcolor}

\begin{document}
	
	\preprint{APS/123-QED}
	
	\title{Boomerang quantum walks}

    \affiliation{QuIIN - Quantum Industrial Innovation, EMBRAPII CIMATEC Competence Center in Quantum Technologies, SENAI CIMATEC, Av. Orlando Gomes 1845, 41650-010, Salvador, BA, Brazil.}

    \affiliation{Laborat\' orio de F\'isica Te\' orica e Computacional, Departamento de F\'isica, Universidade Federal de Pernambuco, 50670-901 Recife, PE, Brazil.}

    \affiliation{Instituto de F\'isica, Universidade Federal de Alagoas, 57072-900 Macei\'o, Alagoas, Brazil.}

    \author{A. R. C. Buarque$^{1,2}$}
    \author{W. S. Dias$^{3}$}
    \author{Ernesto P. Raposo$^{2}$}

\begin{abstract}

In this study, we investigate the emergence of the quantum boomerang effect in discrete-time quantum walks (DTQWs) subjected to random phase disorder. Our analysis shows that this effect can arise solely from the intrinsic momentum dynamics of the DTQW, without requiring external bias or asymmetry. We explore the evolution of the mean position of the quantum walker, denoted as $\overline{X}(t)$, under various initial conditions of the walker and quantum coin operators. The results indicate a significant dependence of the observed phenomena on the choice of initial state, enabling the selective induction of the quantum boomerang effect in both or only one portion of the wavepacket associated with specific internal states. By varying the quantum coin parameter $\theta$, we find that the maximum mean position follows a power-law decay near the Pauli-Z coin, characterized by $X_{\text{Max}} \sim \theta^{-2}$. Additionally, we identify a scaling behavior $X_{\text{Max}} \sim W^{-2}$, which is consistent with the localization length observed in disordered quantum systems.
Such a selective nature of the boomerang effect related to internal states reveals valuable insights for controlling quantum transport, which could lead to applications in quantum state management, spatial separation of quantum information carriers, and targeted information retrieval.

\end{abstract}
	\maketitle
	
\section{Introduction}

Understanding the mechanisms that govern information propagation in disordered quantum systems is crucial for describing fundamental quantum phenomena, such as Anderson localization. Originally predicted by P. W. Anderson in 1958~\cite{PhysRev.109.1492}, this effect describes how uncorrelated static disorder can inhibit the diffusion of quantum particles, leading to their spatial localization and thus preventing the coherent transport of information. The implications of Anderson localization extend across various fields, such as electronic transport in condensed matter~\cite{roati2008anderson, modugno2010anderson}, quantum optics~\cite{wiersma1997localization, segev2013anderson}, quantum walks~\cite{PhysRevB.96.144204, PhysRevE.100.032106}, and non-Hermitian systems~\cite{PhysRevB.107.024202}. In recent years, experimental advances have confirmed the occurrence of Anderson localization across a broad range of physical settings, including optical systems~\cite{segev2013anderson, PhysRevLett.134.046302}, acoustic media~\cite{PhysRevB.78.134206}, and ultracold matter waves~\cite{billy2008direct, PhysRevLett.115.240603, bouyer2009anderson}. These studies provide direct evidence of how disorder fundamentally constrains the dynamics of quantum wavefunctions, reinforcing the universality and significance of localization phenomena in contemporary quantum physics.

Among the recently proposed emergent phenomena that occur in the presence of disorder is the quantum boomerang effect (QBE) \cite{PhysRevA.99.023629}, which arises at the delicate boundary between ballistic propagation and complete localization of the wavefunction~\cite{PhysRevA.99.023629,PhysRevA.102.013303,PhysRevB.106.L060301,PhysRevB.107.094204,PhysRevA.109.063315}. Unlike Anderson localization, where a particle remains confined to a specific region of space, the QBE features an initial phase of propagation followed by a gradual return of the mean position  of the wave packet to its starting position. This behavior illustrates the intricate interplay between quantum interference, disorder, and residual coherence, making its investigation important for understanding phase transitions in disordered quantum systems.

In this context, the QBE has been recently observed experimentally in an implementation of the quantum kicked-rotor model using ultracold bosons~\cite{PhysRevX.12.011035}. Nevertheless, this study demonstrated that the suppression of dynamical localization through stochastic kicking also leads to the disappearance of the boomerang effect. Additionally, both mean-field~\cite{PhysRevB.106.L060301,PhysRevA.102.013303} and many-body interactions~\cite{PhysRevB.107.094204} have been shown to suppress the QBE. These results indicate that the effect is highly sensitive to the strength of interactions: as interactions increase, the return of the wave packet becomes incomplete, leading the system to localize at a finite position distinct from the initial one~\cite{PhysRevA.102.013303}. Interestingly, in a two-component Fermi gas, the QBE is preserved only in the total density profile, while the individual spin components fail to exhibit the effect~\cite{PhysRevA.109.063315}.

Although the QBE has been investigated in diverse quantum systems in recent years, here we aim to deepen the understanding of this phenomenon by exploring scenarios in which quantum particles evolve through an interplay between internal and spatial degrees of freedom. This framework is influenced by tunable interference mechanisms shaped by initial conditions. Examples of systems exhibiting such features include DTQWs, which naturally incorporate internal quantum states (such as spin) and allow for precise control over the dynamics at each step. This framework has emerged as a powerful and versatile tool not only in the context of quantum computing~\cite{PhysRevLett.102.180501,childs2013universal,PhysRevA.81.042330,de2021quantum,verstraete2009quantum} and the development of new quantum algorithms~\cite{qiang2024quantum,ambainis2003quantum}, but also for modeling quantum systems in a highly controllable and tunable fashion~\cite{venegas2012quantum,PhysRevA.105.042216,PhysRevA.108.062206,PhysRevE.107.064139}.

DTQWs have already been successfully employed to study Anderson localization and related interference effects~\cite{PhysRevB.96.144204,PhysRevE.100.032106}. Their experimental implementations have further demonstrated the potential of these frameworks as platforms for investigating fundamental aspects of quantum transport~\cite{PhysRevLett.106.180403,Crespi2013}. The introduction of dynamic disorder into these systems has proven particularly significant, as it can not only preserve but also enhance the degree of entanglement between the walker's internal and spatial degrees of freedom~\cite{PhysRevLett.111.180503,PhysRevA.89.042307,wang2018dynamic}. This directly affects how quantum information is distributed throughout the evolution of the system. Additionally, the choice of a quantum coin plays a crucial role in shaping the dynamics of the walk. By varying the type or parameters of the coin, we can alter both the propagation regime and the system's sensitivity to noise. Recent studies have demonstrated that different coin operators can significantly influence the emergence of extreme events~\cite{PhysRevA.108.062206} and self-focusing regimes~\cite{PhysRevA.103.042213}. These phenomena are highly relevant for understanding the dynamics of complex systems and could be useful for designing robust quantum algorithms.

To our knowledge, the manifestation of QBE in DTQWs has remained unexplored so far~\cite{PhysRevA.99.023629,PhysRevA.102.013303,PhysRevB.106.L060301,PhysRevB.107.094204,PhysRevA.109.063315}. Moreover, while previous studies in the literature typically consider an initial Gaussian wave packet endowed with an external momentum~\cite{PhysRevA.99.023629,PhysRevA.102.013303,PhysRevB.106.L060301,PhysRevB.107.094204,PhysRevA.109.063315}, here we propose an alternative perspective. Instead of explicitly introducing an external driving momentum, we investigate the role of the intrinsic momentum that naturally emerges in DTQWs from the structure of the initial state. This intrinsic momentum arises from the internal (coin) state of the walker, which biases the interference pattern and leads to a net displacement during the early stages of the evolution~\cite{jayakody2023revisiting}. This perspective opens a new route for exploring the QBE, prompting us to examine its behavior under varying coin parameters and disordered conditions.

We investigate numerically how the choice of different initial states, quantum coin operators, and disorder strengths influence the emergence of the QBE in one-dimensional DTQWs subject to random phase fluctuations. 
Our results reveal a peculiar behavior of this phenomenon, manifesting itself in the form of new dynamical regimes not previously described in the literature. 
Indeed, while disorder-free systems cannot display the boomerang effect, here we demonstrate that the QBE emerges in disordered systems for any coin and initial conditions composed of asymmetric states, provided the initial intrinsic momentum overcomes the localization trend due to disorder in the walker's early dynamics.  
However, instead of returning to the starting point, as in typical QBE phenomena, after the reversal the DTQW localizes at a different position on the opposite side of the chain, reflecting the direction of the initial drift. 

Moreover, we show that the typical time for the return of the wave packet in the QBE of DTQWs is rather sensitive to the choice of the quantum coin. 
In fact, for coins approaching the Pauli-X operator, Anderson localization features become predominant, rapidly suppressing coherent transport. 
In contrast, coins close to the Pauli-Z choice delay the onset of destructive interference, allowing for a much longer net displacement of the walker before localization mechanisms take over.

We further present a detailed study of the walker’s maximum displacement prior to reversal as a function of the quantum coin and disorder. 
Our findings reveal that the walker's maximum mean position decays inversely with the square of the disorder strength and localization length, significantly affecting the walker’s dynamics. 
The maximum displacement in QBE is also sensitive to the bias parameter of the coin, highlighting its central role in controlling the propagation regime.

The article is organized as follows. In Sec. II we introduce the model and describe the general formalism. Results and discussion are presented in Sec. III. Lastly, final remarks and conclusions are left to Sec. IV.

\section{Model and formalism}

The boomerang quantum walk consists of a qubit propagating in a one-dimensional phase-disordered chain with $N+1$ sites, in which the positions are discrete and indexed by integers, $n=(-N/2, \dots, -1, 0, 1, \dots, N/2)$. 
The qubit is defined in a two-level space called the coin space, \( \mathcal{H}^\mathcal{C} = \{\ket{R} = (1,0)^T, \ket{L} = (0,1)^T\} \), where \( \ket{R} \) and \( \ket{L} \) represent right and left polarizations, respectively, and the superscript $T$ denotes the transpose. 
Moreover, the position space is defined as $\mathcal{H}^{\mathcal{P}} = \{ \ket{n} \}$. 
The total Hilbert space is the tensor product of the coin and position spaces, $\mathcal{H} = \mathcal{H}^\mathcal{C} \otimes \mathcal{H}^\mathcal{P}$. 
The state of the quantum walker at any discrete time $t~(= 0, 1, 2, ...)$ is a superposition of coin and position states:
\begin{equation}
\ket{\Psi(t)}= \sum_n [a_{n}(t)\ket{R} + b_{n}(t)\ket{L}] \otimes \ket{n},
\label{eq:initial-state}
\end{equation}
where $a_{n}(t)$ and $b_{n}(t)$ are the probability amplitudes of the right and left coin states, respectively, at position $n$ and time $t$. The normalization condition reads $\sum_{n}P_{n}(t)=\sum_{n}[|a_{n}(t)|^2 + |b_{n}(t)|^2] = 1$.   

The system's dynamic evolution is obtained through the time evolution operator $\hat{U}$, $\ket{\Psi(t)}=\hat{U}^{t}\ket{\Psi(0)}$, where $\hat{U}^t=\prod_{t'=1}^t\hat{S}\hat{C}_{t',n}\hat{D},$ with the operators $\hat{S}$, $\hat{C}_{t',n}$, and~$\hat{D}$ defined as follows. 
First, the phase-gain operator is written as
\begin{eqnarray}
\label{eq:phase-operator}
\hat{D}=\sum_{c}\sum_{n}e^{i2\pi\nu}\ket{c}\bra{c}\otimes\ket{n}\bra{n},
\end{eqnarray}
where $c=\{R,L\}$. As indicated in Eq.~(\ref{eq:phase-operator}), a global phase~$2 \pi \nu$ is added to the state of the system upon the action of~$\hat{D}$. 
Here we consider $\nu$ a randomly distributed number in the range $[-W,W]$,  where $W$ represents the width of the disorder. 
We remark that the quantum walk can exhibit diffusive behavior or Anderson localization depending on the nature and strength of disorder \cite{PhysRevLett.106.180403,PhysRevB.96.144204,PhysRevE.107.064139}.
On the other hand, when the phase gain~$\nu$ is set to zero (equivalent to choose $W = 0)$, the system is disorder free and exhibits the uniform quantum walk behavior, with the wavefunction spreading out ballistically.

The system's evolution also depends on the internal and spatial degrees of freedom of the qubit. 
So, to incorporate the internal degree of freedom, a unitary operator~$\hat{C}$, known as the quantum coin, is applied. 
In a more general description, a single-site quantum coin is represented by an arbitrary SU(2) unitary matrix~\cite{PhysRevA.77.032326},  
\begin{eqnarray}
\label{quantum-coin}
    \hat{C}(\theta)&=& \cos\theta\ket{R}\bra{R} + \sin\theta\ket{R}\bra{L} \nonumber \\ 
    & &+\sin\theta\ket{L}\bra{R}-\cos\theta\ket{L}\bra{L}.
\end{eqnarray}
Here we control the angle $\theta \in [0, \pi/2]$, which determines the spatial (right or left) bias of the quantum coin in association with the displacement operator $\hat{S}$ (see below). 
For example, in the case of a fair coin, which selects both possible states $(\ket{R}$ or $\ket{L})$ and directions with equal probability, the choice $\theta=\pi/4$ is adopted (Hadamard coin). 
On the other hand, a~choice in the interval $\theta \in [0, \pi/4)$ ($\theta \in (\pi/4, \pi/2$]) favors the~$\ket{R}$~$(\ket{L})$ state and right (left) direction. 

Lastly, the conditional displacement operator~$\hat{S}$  shifts the discrete position of the walker in the chain, depending on the internal state of the coin. 
Specifically, if the coin is in the $\ket{R}$ state, the walker moves one step to the right, while if it is in the $\ket{L}$ state, it moves one step to the left. 
This operation is expressed as
\begin{eqnarray}
    \label{shift-operator}
\hat{S}=\sum_{n=1}^{N}\ket{R}\bra{R}\otimes |n+1\rangle\langle n| + \ket{L}\bra{L}
    \otimes |n-1\rangle\langle n|.
\end{eqnarray}

In all numerical simulations discussed below, we consider a chain large enough to avoid edge effects. 

\section{Boomerang quantum walk}

Our results were obtained by following the time and spatial evolution of a qubit starting from position $n_{0}$, with initial wavefunction generally expressed by
\begin{eqnarray}
    \label{eq:1}
    \ket{\Psi(t=0)}=\cos \left ( \frac{\alpha}{2} \right ) \ket{R,n_{0}} + e^{i\beta}\sin \left ( \frac{\alpha}{2} \right ) \ket{L,n_{0}}.
\end{eqnarray}
Below we consider the middle of the interval (i.e., the origin $n_0 = 0$) as the walker's starting point.
The initial state~(\ref{eq:1}) depends on two parameters, $\alpha$ and $\beta$, giving us the possibility to explore different states on Bloch sphere. 

We consider three representative initial conditions in Eq.~(\ref{eq:1}), described as follows. 
For (i)~$\alpha = \beta = 0$, the initial state reads ${\ket{\Psi_{R}}(t=0)}=\ket{R,n_{0}}$, while for (ii)~$\alpha=\pi$ and $\beta = 0$ one has $\ket{\Psi_{L}(t=0)}=\ket{L,n_{0}}$. 
In~both cases~(i) and~(ii), we note that the initial state is an eigenstate of the operator $\hat{C}(\theta = 0)$ (Pauli-Z coin).
On the other hand, for the choice (iii)~$\alpha = \beta=\pi/2$ the initial state is symmetric on the internal degrees of freedom: $\ket{\Psi_{S}(t=0)}=1/\sqrt{2}(\ket{R,n_{0}} + i\ket{L,n_{0}})$, being an eigenstate of $\hat{C}(\theta = \pi/2)$ (Pauli-X coin).

Therefore, we notice that each one of these initial states is assigned to an intrinsic initial linear momentum (to the right, left, and null, in the cases (i), (ii), and~(iii), respectively). 
This intrinsic momentum is associated with the initial coin state and biases the interference patterns, leading in the cases~(i) and~(ii) to a net displacement of the walker during the early stages of evolution, as shown below.
\begin{figure}
\centering
\includegraphics[width=0.9\linewidth]{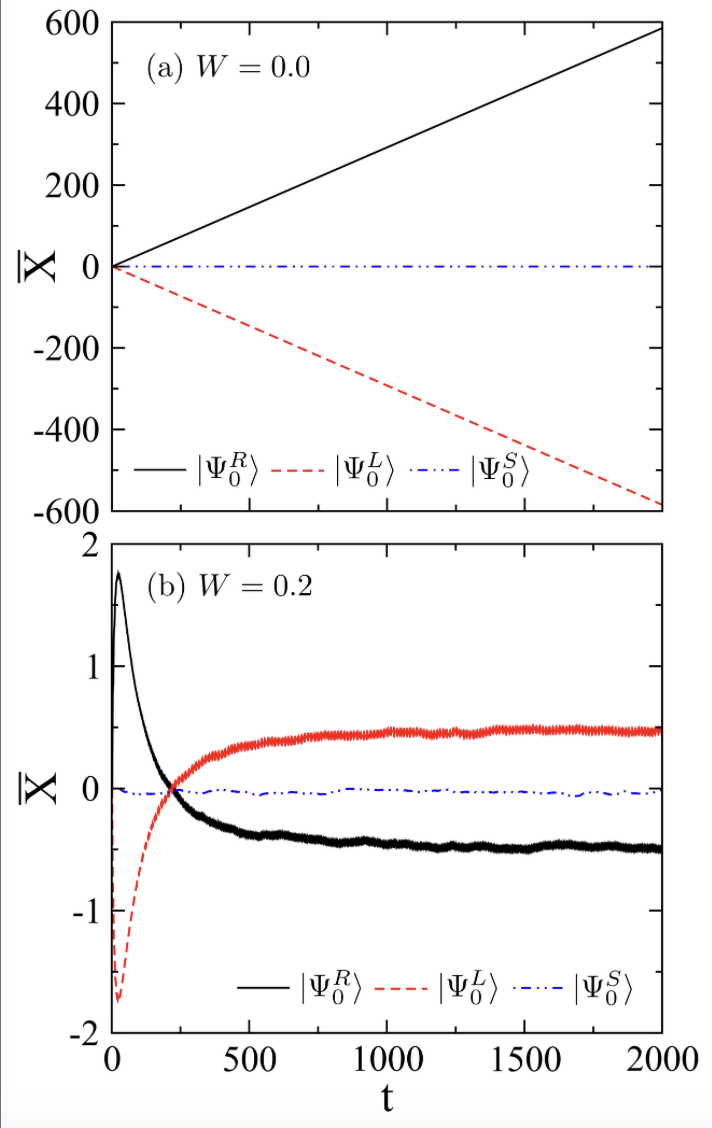}
\caption{Dynamics of the mean position $\overline{X}(t)$ (centroid) of the qubit in the DTQW for the Hadamard coin $(\theta = \pi / 4)$ and three different initial states: ${\ket{\Psi_R(t=0)} \equiv \ket{\Psi_0^R} = \ket{R,n_{0}}}$  (solid~black line) [initial condition~(i), see text]; ${\ket{\Psi_L (t=0)} \equiv \ket{\Psi_0^L} =\ket{L,n_{0}}}$ (dashed red line) [(ii)]; and~$\ket{\Psi_S (t=0)} \equiv \ket{\Psi_0^S} =1/\sqrt{2}(\ket{R,n_{0}} + i\ket{L,n_{0}})$ (blue dotted-dashed line)~[(iii)], with $n_0 = 0$. 
(a)~Disorder-free case $(W=0)$. 
The walker propagates ballistically along the chain for the asymmetric conditions~(i) and~(ii), while for the initially symmetric condition~(iii) $\overline{X}(t)$~remains at the initial position $n_0 = 0$. 
(b)~When disorder is introduced $(W=0.2)$, the dynamics changes significantly for initial conditions~(i) and~(ii). 
The qubit initially exhibits ballistic propagation but quickly returns, evidencing the emergence of the QBE phenomenon. 
The long-term mean position is near---but not exactly at---the starting point:  $\overline{X} \approx -1/2$ [$\overline{X} \approx 1/2$] in the case~(i) [(ii)].
}
\label{figure1}
\end{figure}

We begin by examining the time evolution of the qubit’s wave packet mean position (centroid), which results from an average over 50,000 disorder realizations. The centroid is defined as
\begin{eqnarray}
    \overline{X}(t) = \sum_{n}(n-n_{0})P_{n}(t),
    \label{eq6}
\end{eqnarray}
where  $P_{n}(t) =|\langle \Psi(t) \ket{R,n}|^{2} + |\langle \Psi(t) \ket{L,n}|^{2}$. 
In Fig.~\ref{figure1}, we analyze the dynamics for a Hadamard quantum walk $(\theta=\pi/4)$, considering the three initial states described above: (i) $\ket{\Psi_{R}(0)}$ (solid black line), (ii) $\ket{\Psi_{L}(0)}$ (dashed red line), and (iii) $\ket{\Psi_{S}(0)}$ (blue dotted-dashed line). 
We first track the evolution of the mean position for the disorder-free quantum walk without random phase fluctuations $(W=0)$, as shown in Fig.~\ref{figure1}(a). 
For a uniform DTQW with initial condition~(i), the intrinsic initial momentum causes the probability distribution $P_n(t)$ to develop a more pronounced peak moving to the right, while a small portion of the wave packet shifts to the left (see below). 
This asymmetry results the centroid $\overline{X}(t)$ being displaced to the right. 
A similar behavior occurs for the DTQW with the initial condition~(ii), but in a mirrored way: the more pronounced peak in $P_n(t)$ moves to the left, shifting the centroid in the same direction. 
In contrast, for initial condition~(iii) the probability distribution evolves symmetrically, keeping the centroid at the initial position $n_{0} = 0$. 
No QBE is observed in these disorder-free $(W=0)$ settings.

In remarkable contrast, when disorder is introduced ${(W=0.2)}$, the dynamics changes significantly, see Fig.~\ref{figure1}(b). For the initial condition~(i), the centroid $\overline{X}(t)$ initially shifts to the right but then returns, displaying a clear signature of the QBE behavior. 
However, an interesting dynamics is observed for later times, once the wave packet surpasses the initial position and stabilizes around $\overline{X} \approx -1/2$, as further discussed below. 
The same phenomenon occurs for the initial condition~(ii), showing the boomerang effect to the left of the initial position and subsequent return to $\overline{X} \approx 1/2$. 
For the symmetric initial condition~(iii), the centroid remains nearly constant at~$\overline{X} \approx 0$, with no QBE, similar to the symmetric disorder-free case in Fig.~\ref{figure1}(a).

\begin{figure}
\centering
\includegraphics[width=\linewidth]{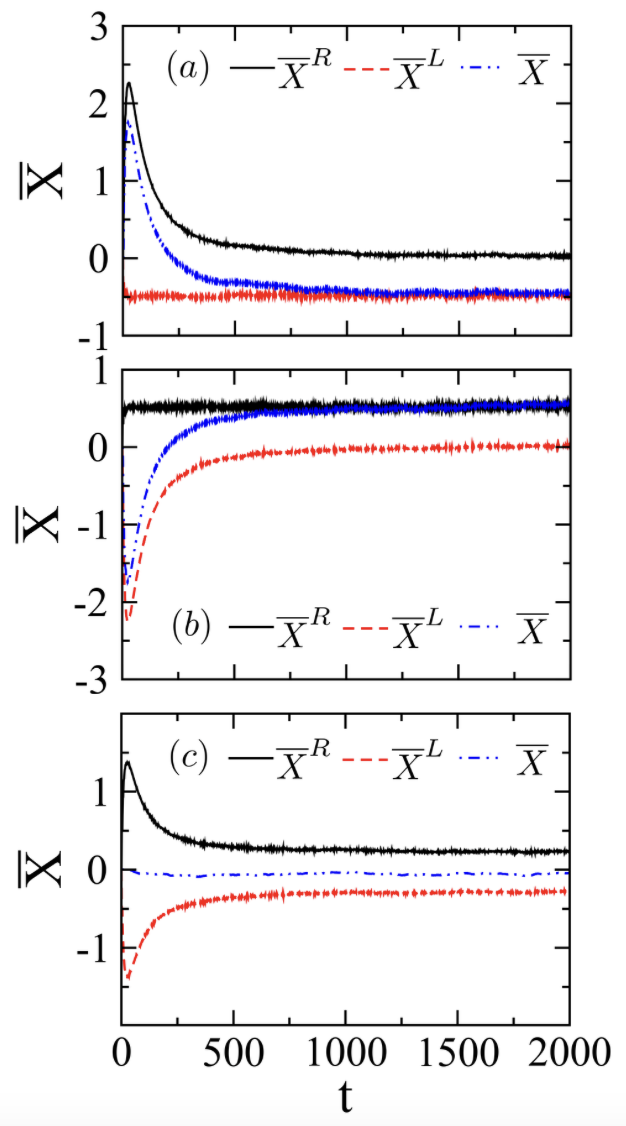}
\caption{Dynamics of the qubit in the DTQW for the Hadamard coin $(\theta = \pi / 4)$, disorder strength $W = 0.2$, and three different initial states: (a) ${\ket{\Psi_R (t=0)}=\ket{R,n_{0}}}$ [initial condition~(i)]; (b) $\ket{\Psi_L (t=0)}=\ket{L,n_{0}}$ [(ii)]; and~(c)~${\ket{\Psi_S (t=0)}=1/\sqrt{2}(\ket{R,n_{0}} + i\ket{L,n_{0}})}$ [(iii)], with $n_0 = 0$. 
In each case, the time evolution of the parts of the centroid associated with each probability distribution is displayed: $\overline{X}^{R}(t)$ associated with $P_{n}^{R}(t)$ (solid black line), $\overline{X}^{L}(t)$ with $P_{n}^{L}(t)$ (dashed red line), and the total displacement $\overline{X}(t)$ associated with $P_{n}(t)$ (dotted-dashed blue line).
The QBE behavior is evidenced in $P_{n}(t)$ for initial conditions~(i) and~(ii), respectively in~(a) and~(b). 
}
\label{figure2}
\end{figure}

To deepen the analysis of the QBE behavior, we next investigate the role of the components that make up the probability distribution $P_{n}(t)$. 
By writing $P_{n}(t) = P_{n}^{R}(t) + P_{n}^{L}(t)$, where $P_{n}^{R}(t) = | \bra{R,n} \Psi(t) \rangle |^{2}$ and $P_{n}^{L}(t) = | \bra{L,n} \Psi(t) \rangle|^{2}$, we quantify the influence of each component of the coin basis states $\{\ket{R},\ket{L}\}$ on the walker's dynamics.
Figure~\ref{figure2} displays the time evolution of the parts of the centroid~(\ref{eq6}) associated with each probability distribution: $\overline{X}^{R}(t)$ associated with $P_{n}^{R}(t)$ (solid black line), $\overline{X}^{L}(t)$ with $P_{n}^{L}(t)$ (dashed red line), and the total displacement~$\overline{X}(t)$ associated with $P_{n}(t)$ (dotted-dashed blue line). 
We have considered the same configurations as in Fig.~\ref{figure1}(b), that is, Hadamard quantum coin $(\theta = \pi/4)$ and disorder strength $W=0.2$. 
The plot is divided into three initial settings: in (a) the qubit starts fully polarized to the right [initial condition (i)]; in (b), fully polarized to the left [(ii)]; and in (c), a symmetric superposition of both polarization states~[(iii)].

\begin{figure}
\centering
\includegraphics[width=0.8\linewidth]{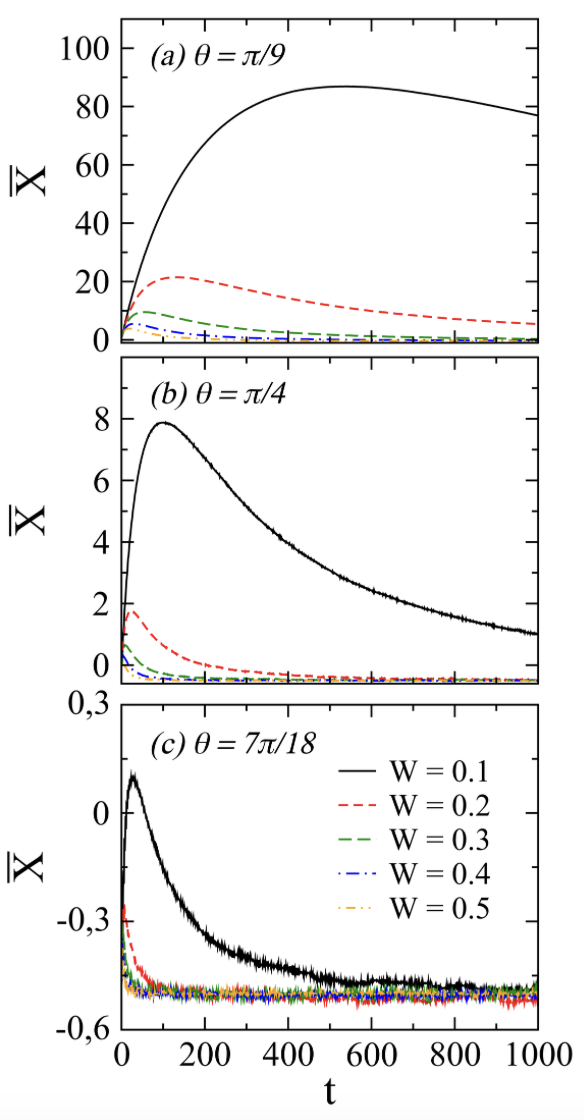}
\caption{Dynamics of the centroid $\overline{X}(t)$ of the qubit in the DTQW for the initial condition~(i) ($\ket{\Psi_R (t=0)}=\ket{R,n_{0}}$, with $n_0 = 0$) and 
disorder strengths $W = 0.1$ (solid black line), $0.2$ (short-dashed red line), $0.3$ (dashed green line), $0.4$ (dash-dotted blue line), and $0.5$ (double-dotted dashed purple line). 
Results for three different quantum coins are shown. 
(a)~The~$\theta = \pi/9$~coin favors state $\ket{R}$ and spatial bias to the right. 
(b)~The Hadamard coin $(\theta = \pi / 4)$ selects the two possible coin states and directions with equal probability. 
(c)~The~$\theta = 7\pi/18$~coin favors state $\ket{L}$ and spatial bias to the left.
Plots~(a) and~(b) display the QBE behavior for $W \in [0.1, 0.5]$, though it is nearly vanished for strong disorder $(W = 0.5)$. 
In~(c) the QBE is noticed only in the weakly disordered regime $(W = 0.1)$.
}
\label{figure3}
\end{figure}

In Fig.~\ref{figure2}(a), we observe that $P_{n}^{R}$ mainly drives the dynamics to the right of the mean position~$\overline{X}(t)$ of the wavefunction for the initial condition~(i). 
Indeed, we note that $\overline{X}^R(t)$ associated with the states $\{\ket{R,n}\}$, which dominate the intrinsic momentum of the DTQW in this case, tends to return to the initial position $n_{0} = 0$, exhibiting QBE behavior. 
However, the intrinsic momentum related to the states $\{\ket{L,n}\}$, which emerges over time due to the application of the quantum coin, does not exhibit the boomerang effect, but remains localized around a finite position $\overline{X}^{L} \approx -1/2$, showing a behavior similar to that expected in the usual case~\cite{PhysRevA.99.023629}.
This explains why, when analyzing the total displacement, the long-term mean position $\overline{X}(t)$ does not tend to return to the starting point $n_0 = 0$ in the~QBE with initial condition~(i), but instead approaches $\overline{X} \approx -1/2$. 

An analogous pattern occurs for the initial condition~(ii), but in a mirrored way, see Fig.~\ref{figure2}(b). 
In Fig. \ref{figure2}(c), we present the case of a DTQW with symmetric initial condition~(iii). 
Observing the evolution of the mean position associated with the probability densities for the states $\{\ket{R,n}\}$ and $\{\ket{L,n}\}$, we see that they exhibit a symmetric dynamics, causing $\overline{X}(t)$ to remain nearly constant at the initial position.

Since the boomerang effect is highly dependent on interference effects throughout the temporal evolution, we next investigate how different quantum coins influence its manifestation under varying degrees of disorder.
To this end, we show in Fig.~\ref{figure3} the time evolution for three quantum coin configurations [$\theta = \pi/9$~(a), $\pi/4$~(b), $7\pi/18$~(c)], considering five disorder strengths: $W = 0.1$ (solid black line), $0.2$ (short-dashed red line), $0.3$ (dashed green line), $0.4$ (dash-dotted blue line), and~$0.5$~(double-dotted dashed purple line). 
As the centroid dynamics is mirrored regarding the initial conditions~(i) and~(ii), we focus below only on the evolution with the initial state~(i). 

For the quantum coin with $\theta=\pi/9$, the QBE is observed for all disorder strengths in the range $W \in [0.1, 0.5]$, as shown in Fig.~\ref{figure3}(a).
We notice in the weak disorder regime $(W = 0.1)$ that the qubit undergoes a significant displacement to the right, reaching a maximum mean position $\overline{X}_{\text{max}} \approx 90$ before starting a slow return. 
As the disorder strength~$W$ is increased, the boomerang effect remains present up to $W = 0.5$, but the maximum displacement decreases drastically, e.g., $\overline{X}_{\text{max}} \approx 5$ for $W = 0.5$. 
Higher values of $W$ lead the maximum mean position to approach the initial position, $\overline{X}_{\text{max}} \rightarrow n_0 = 0$, destroying the QBE behavior, as the initial intrinsic momentum cannot overcome the localization trend due to disorder in the walker's early dynamics (see also below).

We further observe in Fig.~\ref{figure3}(a) that the quantum interference process responsible for the qubit’s return takes longer to occur, when compared to the Hadamard case $(\theta = \pi/4)$ for the same~$W$, shown in Fig.~\ref{figure3}(b), since the momentum distribution of the quantum walk for $\theta = \pi / 9$ contains higher velocity components to the right. 
We recall that, at each time step, the Hadamard coin selects the possible coin states $(\ket{R}$ or $\ket{L})$ with equal probability, while the coin with $\theta = \pi/9 \in [0, \pi/4)$ favors the state $\ket{R}$.
This mechanism causes the probability distribution for $\theta = \pi/9$ to spread more rapidly to the right before destructive interference begins to act significantly. 
As a consequence, at the initial stages of evolution the probability density is more dispersed at the edges of the distribution, leading to a greater maximum displacement before the return occurs. 
Thus, the boomerang effect for $\theta = \pi / 9$ takes place more gradually and over longer timescales, in contrast to the case of the Hadamard coin $(\theta = \pi/4)$, where destructive interference is established more quickly due to the symmetric structure of the probability distribution and the balanced coupling between the coin basis states $\{\ket{R}, \ket{L}\}$.

\begin{figure}
\centering
\includegraphics[width=\linewidth]{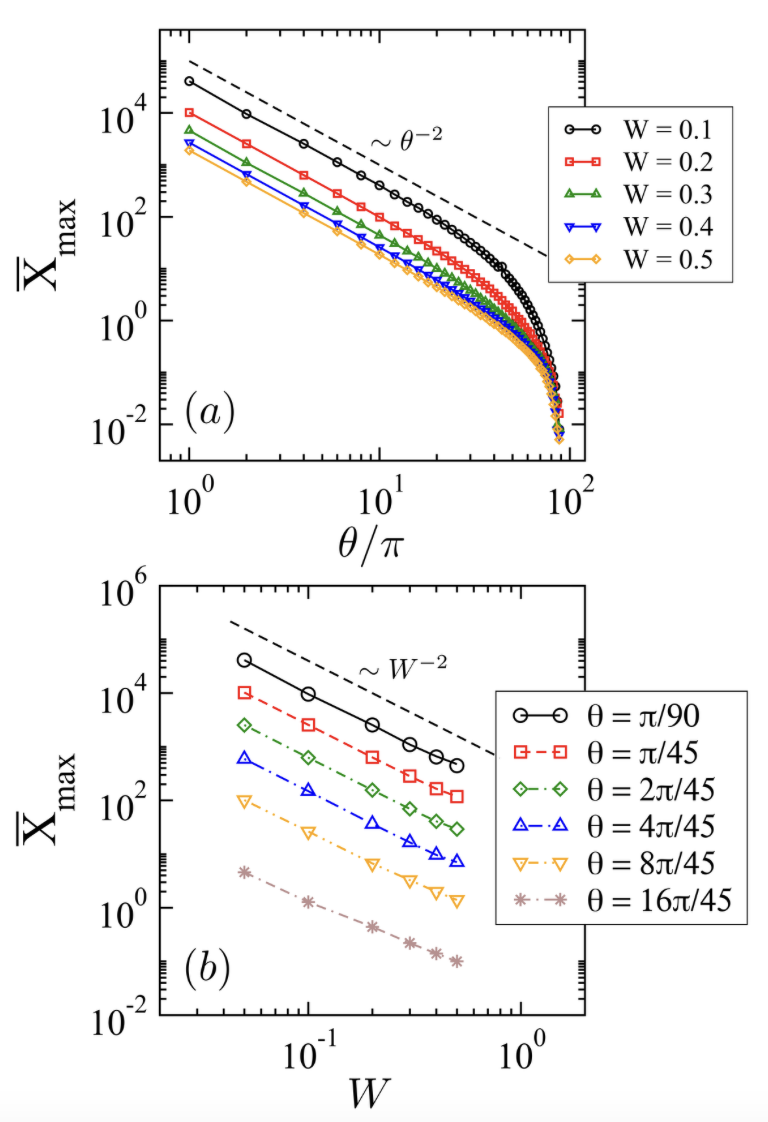}
\caption{Maximum mean position $\overline{X}_{\text{max}}$ of the qubit in the DTQW for the initial condition~(i) ($\ket{\Psi_R (t=0)}=\ket{R,n_{0}}$, with~$n_0 = 0$), (a)~as a function of the coin parameter $\theta \in (0, \pi/2]$, for disorder strength $W \in [0.1., 0.5]$, and~(b)~as~a~function of $W \in [0.05, 0.5]$, for $\theta \in [\pi/90, 16\pi/45]$. 
We notice the power-law decays (a) $\overline{X}_{\text{max}} \sim \theta^{-2}$, $\theta \to 0$ (near Pauli-Z coin), and~(b)~${\overline{X}_{\text{max}} \sim W^{-2}}$. 
The $\theta^{-2}$ decay in~(a) breaks down close to the Pauli-X coin $(\theta = \pi / 2)$.
}
\label{figure4}
\end{figure}

In Fig.~\ref{figure3}(c) we set $\theta = 7\pi/18~\in (\pi/4, \pi/2]$, a coin parameter that favors state $\ket{L}$ and spatial bias to the left. 
In this case, the boomerang effect still occurs for $W = 0.1$, but is destroyed for larger~$W$.
In fact, even~for~$W = 0.2$ the qubit’s mean position remains close to the initial position for all times, not displaying the return~pattern typical of the QBE. 
This behavior suggests that for this value of $\theta$ the destructive interference induced by disorder acts more strongly, significantly limiting the initial propagation of the qubit. 

These findings indicate that the quantum coin also modifies the qubit’s group velocity and its momentum distribution, which directly affect the mean distance traveled by the walker before the return. 
For larger values of $\theta$ with initial condition~(i), the reduced group velocity results in smaller maximum displacements to the right and faster return. 
On the other hand, when the disorder becomes sufficiently strong the qubit’s propagation is drastically suppressed and the QBE can be completely eliminated.




Finally, in order to offer a complementary analysis, we now evaluate the dependence of the wave packet's maximum mean position, $\overline{X}_{\text{max}}$, on different quantum coins in the range $\theta \in (0,\pi/2]$, for various disorder strengths, $W \in [0.1,0.5]$, and with initial condition~(i), see Fig.~\ref{figure4}(a).  
For values of~$\theta$ close to the Pauli-Z coin (i.e.,~for~$\theta \rightarrow 0$), Fig.~\ref{figure4}(a) shows that $\overline{X}_{\text{max}}$ exhibits a decay nearly proportional to $\theta^{-2}$, regardless the disorder strength~$W$. 
For small values of~$W$, the maximum mean position $\overline{X}_{\text{max}}$ near Pauli-Z coin is much larger than the starting position, $n_0 = 0$, signaling the presence of QBE behavior.  
This $\theta^{-2}$ decay observed for small disorder strengths $W$ suggests that, in the weakly disordered regime, the walker’s propagation still follows the standard ballistic behavior governed by the coin parameter \cite{PhysRevA.76.022316}, before strong disorder effects start to dominate, leading to localization for larger~$W$.
Indeed, as~$W$~is enhanced Fig.~\ref{figure4}(a) shows that the value of $\overline{X}_{\text{max}}$ decreases considerably towards~$n_0$, highlighting the role of disorder in suppressing the propagation of the quantum walker.  
A different trend is observed in Fig.~\ref{figure4}(a) for larger values of $\theta$ approaching the Pauli-X coin $(\theta \rightarrow \pi/2)$. 
In~this case, the maximum mean position drops drastically for any~$W$, ceasing to exhibit the $\theta^{-2}$~decay. 
This indicates that the qubit dynamics becomes predominantly influenced by disorder-induced destructive interference, destroying the QBE behavior, consistent with the above discussion. 

Conversely, Fig.~\ref{figure4}(b) presents the behavior of the maximum mean position, $\overline{X}_{\text{max}}$, as a function of the disorder strength $W$, for different quantum coins~$\theta$, and initial condition~(i). 
A clear power-law decay of the form $\overline{X}_{\text{max}} \sim W^{-2}$ is noticed for $W \in [0.05, 0.5]$, regardless the value of $\theta \in [\pi / 90, 16 \pi / 45]$. 
We also observe that this behavior extends up to very low values of $\overline{X}_{\text{max}}$ for large~$W$, a strongly disordered regime in which the QBE phenomenon no longer emerges, in agreement with the scenario displayed in Fig.~\ref{figure3}. Highlighting as predicted numerically that in the vicinity of the Pauli-Z coin, the walker’s spread is inversely proportional to the square of the coin parameter, revealing a fundamental relationship between quantum coin bias and dynamical suppression due to disorder.

\section{final remarks and conclusions}
In this work, we have reported the first evidence of the quantum boomerang effect (QBE) in discrete time quantum walks (DTQWs). 
By analyzing the influence of distinct initial states, quantum coins, and disorder strengths, we have found that the QBE can arise as a consequence of the intrinsic momentum of DTQWs. Our numerical results unveil a previously unreported dynamical regime, in which the quantum walker localizes on the opposite side of the chain, rather than returning to the starting point.  This reversal of the initial drift is observed for all asymmetric initial states, and occurs regardless of the coin operator employed, being eventually destroyed in the strongly disordered regime. 

We have found that the specific choice of the quantum coin determines the timescale at which the QBE becomes prominent. For coins approaching the Pauli-X operator, the walker exhibits strong Anderson localization, with coherent transport being rapidly suppressed by the disorder.  In contrast, using coins closer to the Pauli-Z operator results in delayed localization, which allows for longer-range propagation prior to interference-induced reversal. The selective control of the boomerang effect via internal states reported here provides a promising pathway for state-dependent manipulation of quantum transport. This would enable not only spatial separation of quantum information carriers but also selective information retrieval and tailored quantum dynamics in engineered systems. The sensitivity to internal parameters, such as the coin operator, opens new perspectives for studying non-equilibrium dynamics, symmetry breaking, and localization phenomena in discrete quantum systems.
Furthermore, these results may have implications for the design and control of quantum transport in engineered platforms, such as photonic lattices~\cite{grafe2016integrated,PhysRevLett.100.170506}, trapped ions~\cite{PhysRevA.65.032310}, and superconducting circuits~\cite{PhysRevB.95.144506,flurin2017observing}, where DTQWs are experimentally accessible. 
In particular, the ability to tune dynamical regimes via the coin parameter and disorder strength offers a valuable degree of control for quantum simulations~\cite{PhysRevA.89.032322}, quantum state engineering~\cite{verstraete2009quantum,PhysRevA.109.032229,PhysRevA.96.062326}, and potentially also for robust quantum information protocols in noisy environments.

Future directions can include investigating higher-dimensional extensions, effects of correlated or time-dependent disorder, and the role of measurements and decoherence on the robustness of the QBE. 
Such studies can contribute to a broader understanding of localization dynamics in open quantum systems and their applications in emergent quantum technologies.

\section{Acknowledgments}
This work was partially supported by the Brazilian agencies CNPq (Conselho Nacional de Desenvolvimento Científico e Tecnológico), FAPEAL (Fundação de Apoio à Pesquisa do Estado de Alagoas), and FACEPE (Fundação de Amparo à Ciência e Tecnologia do Estado de Pernambuco).
	
\bibliography{references.bib}

\end{document}